\documentclass[10pt,oneside,onecolumn,a4paper,final]{article}

\usepackage[utf8]{inputenc}
\usepackage{amsmath, amssymb, amsthm}
\usepackage{mathtools}
\usepackage{tikz-cd}
\usepackage{euler}
\usetikzlibrary{arrows}
\usepackage{quiver}
\usepackage{url}

\DeclareFontShape{OT1}{cmr}{m}{up}
      {<->ssub*cmr/m/n}{}

\newcommand{\Alpha}{\mathsf{A}}
\newcommand{\powerset}[1]{\mathbb{P}(#1)}

\title{Modelling Trust and Trusted Systems: A Category Theoretic Approach}

\author{Ian Oliver\\
   \texttt{ian.oliver@oulu.fi}\\
Biomimetics and Intelligent Systems Group\\University of Oulu, Finland
\and 
Pekka Kuure\\ 
 \texttt{pekka.kuure@ndu.fi}\\
National Defence University\\Helsinki, Finland}

\date{\today}

\newcommand{\otrust}{\textbf{TRUST}}

\begin{document}

\maketitle

\begin{abstract}
    We introduces a category-theoretic framework for modelling trust as applied to trusted computation systems and remote attestation. By formalizing elements, claims, results, and decisions as objects within a category, and the processes of attestation, verification, and decision-making as morphisms, the framework provides a rigorous approach to understanding trust establishment and provides a well-defined semantics for terms such as `trustworthiness' and 'justification'/forensics. The trust decision space is formalized using a Heyting Algebra, allowing nuanced trust levels that extend beyond binary trusted/untrusted states. We then present additional structures and in particular utilise exponentiation in a category theoretic sense to define compositions of attestation operations and provide the basis of a measurement for the expressibility of an attestation environment. We present a number of worked examples including boot-run-shutdown sequences, Evil Maid attacks and the specification of an attestation environment based upon this model. We then address challenges in modelling dynamic and larger systems made of multiple compositions.
\end{abstract}

\section{Introduction}

Trusted computing, confidential computing, supply-chain security, run-time attestation etc are technologies that are either built upon or require a form of trusted computing base to operate. This trusted computing base provides guarantees about the identity and integrity of software, hardware, and more typically firmware running those systems \cite{Lipner2015TheBA,nibaldi1979proposed}.

The `modern day' concept of trusted computing revolves around ensuring that a given system, platform, computer, device, cloud \cite{ott2023universal} etc. can provide evidence \cite{ferro2024standard} - ideally cryptographically signed and bound to that device - through a process of attestation and verification, after which a decision can be made on whether that \textit{element} is trusted. This usually involves the use of a Trusted Platform Module \cite{arthur2015practical} and a remote attestation server of which there are numerous examples\footnote{RedHat Keylime, Jane Attestation Server etc.}.

The notion of trust however has deep philosophical roots with the concept taking on many forms from pure philosophical reasonings \cite{simpson2023trust} to intuitionistic logics \cite{kripke1965semantical,moschovakis1999intuitionistic} and on to social and systems theory \cite{luhmann2013introduction,gambetta2000can} and computer science \cite{thompson1984reflections}.

One particular issue is the difference between \textit{trusted} and \textit{trustworthy} \cite{hardin2002trust,11078841,kumar2020evolution}. The latter is often used as a synonym for the former with, for example, statements such as \textit{trustworthy AI} being mistaken as demonstration of the veracity of that AI rather than it being a statement that something is capable of presenting proof of being trusted.

Frameworks for trust are varied, such as those for social and agentic interaction systems \cite{guha2004propagation,primiero2017trust}, for establishing compliance and metrics for trust \cite{shaikh2012trust,ali2022maturity,yan2016security} and for invalidating security properties \cite{cherkaoui2025categorical}. A more general model based on the structures that have evolved through developments in remote attestation is largely missing. Such models are important \cite{10.1007/978-3-030-23182-8_10}, especially in standardisation, where common terminology and understanding are critical. 

The IETF Remote ATtestation procedureS - RATS - standard \cite{rfc9334} currently serves as the canonical definition of the remote attestation process from a technical perspective but addresses this from a data-flow or protocol point of view rather than providing a semantic grounding of this flow. RATS (and also the related TEEP and SEAM standards) do not address what is meant by evidence, expectations, appraisal policies or decisions, nor the relationship between these, other than stating they `exist'. Similarly edge cases such as where some evidence does not exist or is incomplete are not addressed - it is just assumed that all necessary evidence is available and then the (somewhat ignored) decision process is a simple binary yes/no.

In this paper we set out such a framework for establishing a logic of the overall attestation process and formal meanings for objects that make up trusted systems rather than focus on the mechanics (the how of attestation) and also to address the trust decision process from any results obtained from attestation. 

In order to do this we utilise category theory \cite{spivak2014category,goldblatt2014topoi,fong2019invitation} which provides us with a simple modelling language, clear rules of compositionality mechanisms for reasoning  \cite{mabrok2017category,joque2020deconstructing} and the ability to abstract and decompose as necessary. Category theory allows us to more effectively model the structural similarities between domains, eg: Luhmann's and the DoD's as referenced above and bridge between logical, social and other views. We utilise a topos in particular as this facilitates reasoning about trust in practical systems by formalizing the decision-making process within a logical framework.

UML, description logics \cite{oliver2009proposed} etc and introduce `baggage' such as language semantics close to programming language, functions and relationships are typically not first-class etc \cite{8401590,diskin2014category}. 

We continue by first presenting the category \otrust{} which is used to define the core concepts and relationships such as attest, verify, trustworthiness and how these are mapped to the oft neglected decision process. Once the basic model is established when the present a number of additional properties of the model and then continue with a number of worked examples.

The focus of this work is to develop a working model and demonstrate advantages of a category theory approach. We have identified areas for further refinement and formalisation that would benefit from deeper exploration.

\section{The Category \otrust{}}\label{basicstructures}
In this section we present the structure of our category and explain the attestation pipeline and decision mechanism.

\subsection{Basic Structure of the Category}

\otrust{} is equipped with Elements, Claims, Results and Decision objects denoted $E$, $C$, $R$ and $D$. Morphisms $attest$, $verify$ and $decide$ relate these objects as shown in figure \ref{basicobjectsmorphisms} which defines the trust establishment process \cite{datta2009logic}.  

Elements refer to machines, hardware components, software components and potentially complex systems composed of multiple parts \cite{booth20225g}, but here we treat an element as a black-box; we consider composition at a later stage. Claims are bodies of information or evidence about the integrity and identity of the elements, such as Trusted Platform Module - TPM - Quotes, UEFI eventlogs etc, obtained through the process of attestation ($attest$). Results are the outcomes of the verification process ($verify$) where claims are checked against known criteria. Finally, Decisions are the result of a process ($decide$) where the type of result is mapped by some policy to a `trusted' or 'untrusted' state denoting whether we trust or not the element in question. As we shall see the decision space itself may be more nuanced to better describe the differing kinds of trust/security-levels we see in real-life situations.

The initial object $0$ and morphism $0_E$ denote an instance of an element in some state where it can not be attested, for example, before or during  the boot process of a computer. 

The claim, result and decision space carry more structure that enables deeper reasoning about trust and will be presented later.

\begin{figure}[ht]
\centering
\[\begin{tikzcd}
	0 && E && C && R && D
	\arrow["{0_E}", from=1-1, to=1-3]
	\arrow["attest", from=1-3, to=1-5]
	\arrow["verify", from=1-5, to=1-7]
	\arrow["decide", from=1-7, to=1-9]
\end{tikzcd}\]
\caption{Objects and Morphisms}\label{basicobjectsmorphisms}
\end{figure}

The decision space forms part of the topos construction as the subobject classifier with the characteristic morphism $\mathcal{A}:E\rightarrow D$, where  $\mathcal{A}=decide\circ verify\circ attest$ - the attestation pipeline - that classifies elements by their degree of trust. 

\begin{figure}[ht]
\centering
\begin{tikzcd}
	T && 1 \\
	\\
	E && D
	\arrow["{!_T}", from=1-1, to=1-3]
	\arrow["{i_E}", hook, from=1-1, to=3-1]
	\arrow["true", from=1-3, to=3-3]
	\arrow["{\mathcal{A}}", from=3-1, to=3-3]
\end{tikzcd}
\caption{A Pullback showing $T$ are `trusted' parts of $E$}
\label{msubobjectclassifierA}
\end{figure}
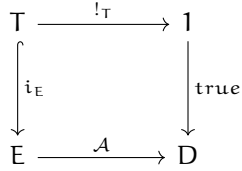

The pullback $T$ denotes elements that are \textit{trusted} as shown in figure \ref{msubobjectclassifierA}. We have more to say on $\mathcal{A}$ and the decision space later. For the moment however this provides a constraint over and defines the logic of the attestation pipeline.

We now explore this basic, abstract model in the following sections.

\subsection{Attestation}
The process of attestation itself is the gathering of individual claims from an element over time. We describe the $attest$ morphism and how it arises and the structures it operates over.

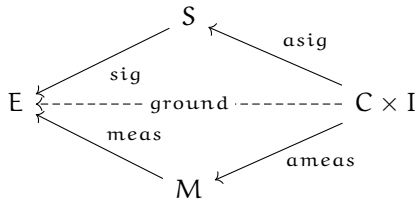
\begin{figure}[ht]
\centering
\[\begin{tikzcd}
	&&& S \\
	& E &&&& {C\times I} \\
	{} &&& M
	\arrow["sig", from=1-4, to=2-2]
	\arrow["asig"', from=2-6, to=1-4]
	\arrow["ground"{description}, dashed, from=2-6, to=2-2]
	\arrow["ameas", from=2-6, to=3-4]
	\arrow["meas"', from=3-4, to=2-2]
\end{tikzcd}\]
\caption{Pullback of Claims}\label{pullbackclaims}
\end{figure}

The relationship between claims and the element can be expressed through a pullback indexed by a `claim identifier' $C\cong (M\times_{E}S)\times I$ as in figure \ref{pullbackclaims}. 

Objects $S$ denote cryptographic signatures (of claim contents) or cryptographic tokens of some form. Objects $M$ denote integrity measurements such as those generated by TPMs, eg: PCRs, Quotes etc, or log files such as UEFI eventlogs etc. A TPM Quote for example consists of both measurements and a cryptographic signature of those measurements relating it with a specific machine or element.

The object $I$ are identifiers used to ensure each claim is unique. These could be constructed by mechanisms such as UUID4 values, or through timestamping and other means.

The morphisms $asig_i$ and $ameas_i$ are simple projections from individual specific, identified claims to the signatures and measurements contained therein. The morphisms $sig$,$meas$ relate signatures and measurements these back to the source element. The morphism $ground$ mediates these through the pullback and \textit{grounds} the claim to a specific element. A further morphism $aid$ extracts or projects the identifier of a given claim.

The morphism $attest:E\times I \rightarrow C$ describes the process of generating a claim, while the pullback explains its structure and ensures that two distinct claims are associated with distinct elements in $E$. Because of this we additionally assert: $ground\circ attest = Id_E$

We must also consider the cases where signatures or even measurements might not be present. We do this by refining $S$ and $M$ to include notions of `null'-signatures and measurements respectively. We define the null values as initial objects $0_M$ and $0_S$ respectively as in eqn.\ref{claimnullassert}.

\begin{equation}
\begin{aligned}
S & = S_+ \sqcup 0_S \\
M & = M_+ \sqcup 0_M  \\
\end{aligned}\label{claimnullassert}.
\end{equation}

We can interpret claims containing null values as either unsigned measurements $(m, 0_S)$, as plain signatures $(0_M, s)$ such as cryptographic tokens without associated integrity information of an element, or as no information $(0_M, 0_S)$. These also correspond to real-world situations where such information might be malformed or due to some transient errors such as loss of network.

A characteristic morphism is defined $\chi_{NULL}:C\rightarrow D$ where $\chi_{NULL}( 0_M,0_S )\rightarrow\bot$ that asserts claims with null measurements and signatures are correctly decided upon as being of the lowest trust-level in any decision space.

There is no temporal ordering on the construction of a claim implied. However, to be strict the process of $attest$ should be always be atomic, for example as the TPM 2.0's quote operation is implemented.

As an example of the above, we attest two machines $e1$ and $e2$ and obtain their TPM's PCR measurements\footnote{PCR measuresments are not signed by the TPM unlike a TPM quote, hence $0_S$ as the signature} \cite{7092916} giving $((m1,0_S),c1)$ and $((m2,0_S),c2)$. If $m1=m2$ then we can say that $e1$ and $e2$ are running the same firmware and firmware configuration (PCR selection depending). As we do not have signatures, the role of $ground$ becomes critical in that allows us ensure that the two pieces of evidence came from different machines.

If we had taken a TPM quote instead, then this locality becomes stronger. If the $s1$ were a signature derived from the TPM's endorsement key, such as the attestation keys, then given claims then the pullback construction of, eg. $(m1,s1,c1)$ ensures $sig(s1)=meas(m1)=e1$.

\subsection{Verification}
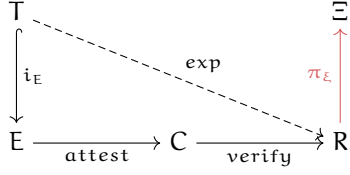
\begin{figure}[ht]
\centering
\begin{tikzcd}[cramped]
	T &&&& \Xi \\
	\\
	E && C && R
	\arrow["{i_E}", hook, from=1-1, to=3-1]
	\arrow["exp", dashed, from=1-1, to=3-5]
	\arrow["attest"', from=3-1, to=3-3]
	\arrow["verify"', from=3-3, to=3-5]
	\arrow["{\pi_\xi}", color={rgb,255:red,214;green,92;blue,92}, from=3-5, to=1-5]
\end{tikzcd}
\caption{Attest and Verify Diagram} \label{attestandverifydiagram}
\end{figure}

The process of verification is described by the morphism $verify:C\times\Xi\rightarrow R$ where resulting structure mirrors $C$ and is augmented by the context as $R\cong(M\times S\times \Xi)$. This is shown in diagram in figure \ref{attestandverifydiagram} which extends that in \ref{basicobjectsmorphisms}.

The object $\Xi$ series as a placeholder for contextual information such as external state, timestamps, metadata, policies, rules and constraints that are not a formal part of the attestation logic. By leaving $\Xi$ intentionally un(der)-defined the model remains general and allows us to focus on the attestation process overall.

The result space is partitioned into a set of spaces in order to capture different kinds of results which are rarely yes-no type answers but also include error conditions (syntax, rule application, network/infrastructure failures), but also with weaker forms of claims. At least two co-products must be defined: a success space and an error space $R_{ERR}$, along with the required injection morphisms $v_n$ which must be monic. The conditions in $verify$ must be mutually exclusive and exhaustive which is obvious in eqn.\ref{verifymorphism} but should be proven for more complex cases as in the proof below.

\begin{equation}
R=\coprod_{1\ldots n} \{R_1...R_n\} \amalg \{ R_{ERR} \}
\end{equation}

The projection $\pi_\Xi:R\rightarrow\Xi$ ensures that the contextual information is available during the verification process. We abuse notation a little and a statement such as $\pi_\Xi(x)$ \textit{is true}, means that the property evaluated in $\Xi$ aligns with the verification process.

\begin{equation}
verify(c,x) = \begin{cases} r_n & \text{if $c$ match an expectation in $T$, and $\pi_\Xi(x)$ is true}  \\
         R_{ERR} & \text{otherwise} \end{cases} 
\label{verifymorphism}
\end{equation}

\begin{proof}
    The morphism \(\text{verify}: C \times \Xi \to R\) is defined for all claims \(C\) and contextual information \(\Xi\). The result space \(R\) is a coproduct of success and error spaces. The verification process is exhaustive and mutually exclusive, as defined in eqn.\ref{verifymorphism}. For any claim \(c\) and context \(\xi\), \(\text{verify}(c, \xi)\) maps to either a success space \(R_n\) or the error space \(R_{ERR}\). Thus, \(\text{verify}\) is total.
\end{proof}

We introduce 3 characteristic morphisms $\chi_S:S\rightarrow \{\top,\bot\}$ and $\chi_M$ and $\chi_I$ similarly. The former checks that a valid signature is present and that any payload is verified against this, the second check that the measurement is the one expected and the latter checks for the freshness of the claim. In TPM Quote terms: the measurements are signed by a valid attestation key, the measurements are expected and finally freshness though timestamp and nonce validation.

Since $\mathcal{A}=decide\circ verify\circ decide$ and $T\cong\{e\in E\;|\;\mathcal(A)=\top\}$ then $T$ is the space of all elements that are running ``correct" firmware and have provided valid and fresh cryptographic proof of doing so. The statement $exp = i_E\circ attest\circ verify$ commutes thereby picking those elements that are in the ideal or optimal state.

At this point we can start to present an example. We define 3 success classes ${1,S,M}$ denoting full success, valid signature only and valid measurement only respectively. The definition of $R$ in eqn.\ref{rexample} and case statement for our worked example of $verify$ is given in eqn.\ref{verifyexample_revised}.

\begin{equation}
R \cong \left( \coprod_{r \in {1, S, M}} r \right) \amalg { R_{ERR} }\label{rexample}
\end{equation}

\begin{equation} verify(c,x) = \begin{cases} 
  v_1 & \text{if } \chi_S(asig(c)) \wedge \chi_M(ameas(c)) \wedge \chi_I(aid(c)) \wedge \pi_\xi(x) \\  
  v_S & \text{if } \chi_S(asig(c)) \wedge \chi_I(aid(c)) \neg \chi_M(ameas(c)) \\ 
  v_M & \text{if } \neg \chi_S(asig(c))  \wedge \chi_I(aid(c)) \wedge \chi_M(ameas(c)) \\
  v_{ERR} & \text{if } (asig(c), ameas(c)) = (0_S, 0_M) \vee \\
          & \qquad \qquad \qquad\qquad\qquad \neg\pi_\xi(x) \vee \neg \chi_I(aid(c))  \end{cases} 
\label{verifyexample_revised}
\end{equation}

The condition $(asig(c),ameas(c))=(0_S,0_M)$ specifically targets the initial object $0_C$ of the claim space. By mapping the initial claim to the error injection $v_{ERR}$, we preserve the property that a total lack of evidence must result in a total lack of trust.

In summary, $exp$ provide a link to the ideal states, in the verification process itself:
\begin{itemize}
 \item a compromised element fails $\chi_M$
 \item an impersonated element fails $\chi_S$
 \item a replayed or `stale' session fails $\chi_I$ 
\end{itemize}
 \subsection{Decisions}\label{decisions}
 Now we can explore the mechanism by which decisions are made and it is here, bearing in mind that we have other potential cases to consider other than just boolean yes-no answers. The formalisation of $decide$ depends upon what structure and `levels' of trust and the logical formalism inside $D$; in reality it is influenced by the kinds of results and nuances of security which is reflected in the co-products of $R$ and context $\Xi$. 
 
 \subsubsection{The Decision Process}
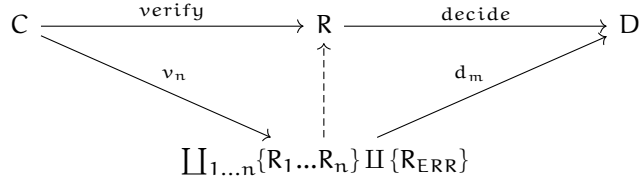
\begin{figure}[ht] 
 \centering 

 
 \[\begin{tikzcd} 	C && R && D \\ 	\\ 	&& {\coprod_{1\ldots n} \{R_1...R_n\} \amalg\{ R_{ERR} \}} 	\arrow["verify", from=1-1, to=1-3] 	\arrow["{v_n}", from=1-1, to=3-3] 	\arrow["decide", from=1-3, to=1-5] 	\arrow[dashed, from=3-3, to=1-3] 	\arrow["{d_m}", from=3-3, to=1-5] \end{tikzcd}\] 
 \caption{Mapping to Decisions}
 \label{mappingtodecisions} 
 \end{figure}

 The total mapping $decide$ is an aggregation or fold of projection functions as illustrated in eqn.\ref{decidefold} and in figure \ref{mappingtodecisions}.  
 
 \begin{equation}   
 decide : R \rightarrow D = [ d_1, d_2, \dots, d_n, d_{ERR} ] 
 \label{decidefold} 
 \end{equation}  
 
 For every injection $v_n:R_n\rightarrow R$ defined in $verify$, there is a corresponding decision morphism $d_m:R_m\rightarrow D$ such that $decide(v_i(r_i), \xi) = d_i(r_i, \xi)$ commutes for all $i$ as in the proof below.

\begin{proof}
    The morphism $decide:R \rightarrow D$ is defined for all results $R$. The decision space $D$, which will be formally defined as a Heyting Algebra later, provides a structured lattice for trust decisions. The morphism $decide$ maps every result in $R$ to a decision in $D$:
    \[
    decide = [d_1, d_2, \ldots, d_n, d_{ERR}]
    \]
    Each injection $v_n: R_n \rightarrow R$ has a corresponding decision morphism $d_n: R_n \rightarrow D$, ensuring that every result in $R$ is mapped to a decision in $D$, therefore $decide$ is total.

    Since both $verify$ and $decide$ are total, their composition $decide \circ verify$ is also total. This ensures that the attestation pipeline $\mathcal{A}$ is total, as it is constructed from total morphisms.
\end{proof}
 
 The decision mechanism is bound by the subobject classifier constraints in the topos: the first is that it must preserve the mapping from $T\rightarrow 1$ and $R_{ERR}$ must map $\bot$ in the decision space to preserve $\chi_{NULL}$.

 
 \subsubsection{Decision Spaces} 
 The choice of formalism for the decision space is critical to how we understand and process trust. As stated simple boolean space may not capture the subltlies of trust and what actions are taken depending upon the trust evaluation. For example, introducing a new device into a system, while it may be trustable, the very fact is it `new' implies that we might not trust it to the highest degree. Similar lattice-like structures appear regularly in other trust domains \cite{10.1007/978-3-540-45217-1_18,coker2008attestation}.  
 
 We consider trust to be intuitionistic in nature \cite{amoore2014security}: lack of evidence does not imply that a system is trusted, or that something not being not trusted is not the same as being trusted \cite{becker2012foundations}. The model is agnostic to the decision space, provided it functions as a subobject classifier.   
 
 To capture this intuitionistic nature, we utilise a Heyting Algebra to encode trust levels \cite{ren2025algebraic} in our decision space. This structure contains a lattice with least and greatest elements $\bot$ and $\top$, relation $\leq$, and is further equipped with operators meet $\wedge$, join $\vee$ and implication $\rightarrow$ such that $(c\wedge a)\leq b$ is equivalent to $c\leq(a\rightarrow b)$. The lack of excluded middle allows us to better reason about the idea of not trusting something: $\neg\neg a\not=a$.  
 
 We present a simple worked example where we define 6 trust levels to capture nuances of the attestation process. These values are presented below and ordering as in eqn.\ref{dsordering}.  
 
 \begin{itemize}     
 \item $\top$ (Top) corresponding to a successfully attested system for identity and integrity
 \item $D_{NEW}$ corresponding to a successfully attested system for identity and integrity, but one which is `new' to our knowledge and need to be monitored in some sense.
 \item $D_{AUTH}$ validated identity without any integrity measurement present     \item $D_M$ validated integrity measurements without any signature present      \item $D_S$ validated identity (signed) but invalid integrity measurements
 \item $\bot$ (Bottom) absolute failure
 \end{itemize}  
 
 The ordering and other properties are: 
 \begin{equation} 
 \begin{aligned}
 & \bot \leq D_S \leq D_{AUTH} \leq D_{NEW} \leq \top \\
 & \bot \leq D_M \leq D_{NEW} \leq \top \\
 & D_{AUTH} || D_M \;\;\; \text{(incomparable) and}\;\;D_{AUTH}\wedge D_M = \bot
 \end{aligned} \label{dsordering} \end{equation}  

If $\mathcal{A}(e)=d$ and $d\neg=\top$, the element is still classified. In this case, $e$ is a member of a more nuanced `intuitionistic' subobject $T_d$ rather than the optimal (trusted) $T$.
 
The incomparability of $D_{AUTH}$ and $D_M$ comes from the handling of the null signature in the pullback $M\times_E S$: neither provides a subset of the other's guarantees they can not be ordered. The implication of this is that in order to reach a higher level of trust, a claim must contain both a measurement and a signature. 
 
 With the justifications that a recognised element with an invalid measurement provides more information than a failed signature. A signature and a null measurement provides more information than one with a failed measurement; $D_{AUTH}$ representing a valid state (authorisation only) whereas $D_S$ represents a possible or known compromise. $D_{AUTH}\leq D_{NEW}$ is because a new element also has to justify its integrity; similarly for $D_{M}\leq D_{NEW}$. Finally $D_{NEW}\leq\top$ with the addition of $\xi$ represents the accumulation of [historical] context  
 
 We can now write a fuller form of $decide$ for our running example and address each of the cases and how these map to $D$. Two projection morphisms are introduced $\pi_\Xi:R\rightarrow\Xi$ and $\pi_C:R\rightarrow C$.  
 
 \begin{equation} 
 \begin{aligned} decide(r) = \begin{cases}
 \top  & \text{if } r \in im(v_1) \wedge \pi_{\Xi}(r)\not=\text{``new''} \\
 D_{NEW}  & \text{if } r \in im(v_1) \wedge \pi_{\Xi}(r)=\text{``new''} \\
 D_{AUTH}  & \text{if } r \in im(v_s) \wedge ameas(\pi_C(r))=0_M \\
 D_S & \text{if } r \in im(v_s) \wedge ameas(\pi_C(r)) \neq 0_M \\
 D_M  & \text{if } r \in im(v_m) \\
 \bot & \text{otherwise } r \in im(v_{ERR}) \\
 \end{cases}  \end{aligned} \label{decide}
 \end{equation}  
 
 We could write the final case as $\bot \text{if } r \in im(v_{ERR})$, but the use of `otherwise' emphasises totality and the mapping to $\bot$, even if we sacrifice some syntactic clarity. The construction in eqn.\ref{decide} maps each co-product in $R$ to a decision unambiguously.

\section{Further Properties}
In this section we extend the basis structures presented in section \ref{basicstructures} to complete our model and allow us to reason about additional features such as how attestation systems may be constructed.

\subsection{Trustworthiness and Judgement}\label{trustjudgement}
So far we have defined the three individual steps that are required to take us from an element to a decision about its trust and the subobject classifier which provides logical constraints over this. The morphism $\mathcal{A}$ can now be better defined as in eqn.\ref{attestationpipeline}. Throughout this process $\Xi$ remains constant.

\begin{equation}
\mathcal{A}:(E\times\Xi)\xrightarrow{attest\times id_\Xi}(C\times\Xi)\xrightarrow{verify}R\xrightarrow{decide}D
\label{attestationpipeline}
\end{equation}

As we shall see later $\mathcal{A}$ can be further refined or internalised as an evaluation over some attestation, verification and decisions functions $\mathcal{A}\cong(-,\tau)$ where $\tau$ is a particular combination of those functions.

The concept of `trustworthiness' can now be given a more precise meaning though it is often mistakenly equated with something being `trusted'. Trustworthiness is the ability to provide evidence so that verification can be performed \cite{10.1145/3567445.3571105}. We define a morphism $\mathcal{W}:E\times\Xi\rightarrow R$ which we call \textit{trustworthy}) where $\mathcal{W}=verify \circ (attest \times id_\Xi)$.

By explicitly defining $\mathcal{W}$ we emphasise the split between evidence gathering - the \textit{how of process} - and the decision process. This is an often misunderstood difference in the overall attestation processes. It also ensures that we can now clearly say that being \textit{trustworthy} does not equate with being \textit{trusted}.

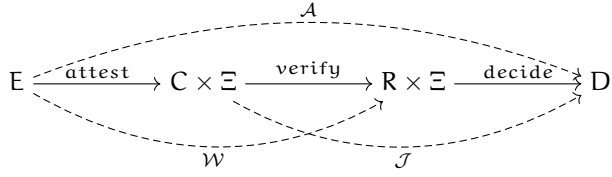
\begin{figure}[ht]
\centering
\begin{tikzcd}[cramped]
	E && {C\times\Xi} && {R\times\Xi} && D
	\arrow["attest", from=1-1, to=1-3]
	\arrow["{\mathcal{W}}"', curve={height=30pt}, dashed, from=1-1, to=1-5]
	\arrow["{\mathcal{A}}", curve={height=-30pt}, dashed, from=1-1, to=1-7]
	\arrow["verify", from=1-3, to=1-5]
	\arrow["{\mathcal{J}}"', curve={height=30pt}, dashed, from=1-3, to=1-7]
	\arrow["decide", from=1-5, to=1-7]
\end{tikzcd}
\caption{Attestation Pipeline, Trustworthy and Judgement}\label{aptj}
\end{figure}

A third morphism which we name ``judgement" can be also constructed in a similar manner $\mathcal{J}:C\times\Xi\rightarrow D$ with definition $\mathcal{J}=decide \circ verify$. These all together can be seen in figure \ref{aptj}.

This morphism is particularly interesting as it can be utilised to provide information about how much evidence is required in order to achieve a given trust level in $D$. If we treat the trust levels in $D$ as propositions, we can utilize the implication $a\rightarrow b$ - $max(x)$ such that $a\wedge x\leq b$ to perform "Gap Analysis":

For example, to answer `what is the minimum evidence (x) required to move from the current trust level to a target level'

The Query ``what is the minimum evidence required to move from the current decision level to a target level: $d_{current}\rightarrow d_{target}$" can be expressed, if $x$ is an element that is classified as $D_{AUTH}$ then the implication $D_{AUTH}\rightarrow\top$ identifies what is missing to reach fully trusted, which in this case are the integrity measurements and establishing that the element is sufficiently proven, ie: not \textit{new}.

The attestation pipeline is defined as:
\[
\mathcal{A} = \text{decide} \circ \text{verify} \circ (\text{attest} \times \text{id}_{\Xi})
\]
where:
\begin{itemize}
    \item $\mathcal{W} = \text{verify} \circ (\text{attest} \times \text{id}_{\Xi})$
    \item $\mathcal{J} = \text{decide} \circ \text{verify}$
\end{itemize}

\begin{proof}
To prove commutativity, we need to show that the composition of these morphisms respects the order of operations and preserves the structure of the pipeline. By definition $\mathcal{A} = \text{decide} \circ \mathcal{W}$. Since $\mathcal{W}$ is the composition of $\text{verify}$ and $\text{attest} \times \text{id}_{\Xi}$, the pipeline is associative:
    \[
    \mathcal{A} = \text{decide} \circ (\text{verify} \circ (\text{attest} \times \text{id}_{\Xi}))
    \]
    This composition is commutative in the sense that the order of operations is preserved, and the pipeline is well-defined.
\end{proof}

\begin{proof}
    $\mathcal{J}$ is defined as $\text{decide} \circ \text{verify}$. Since $\text{decide}$ and $\text{verify}$ are morphisms in the category, their composition is associative and commutative with respect to the pipeline. The pipeline $\mathcal{A}$ is constructed such that:
    \[
    \mathcal{A} = \mathcal{J} \circ (\text{attest} \times \text{id}_{\Xi})
    \]
    This ensures that the pipeline is commutative, as the composition of morphisms respects the order of operations and the structure of the category.
\end{proof}

The consistency of the context $\Xi$ is maintained throughout the pipeline by treating $\mathcal{A}$, $\mathcal{W}$, and $\mathcal{J}$ as morphisms over $\Xi$. The identity morphism $Id_\Xi$ ensures that the environmental parameters used during the extraction of evidence in attest are the same parameters evaluated by the logic in $decide$, preventing context modification attacks during the trust establishment process.

The context $\Xi$ represents external information such as state, timestamps, metadata, policies, and constraints. To prove that $\Xi$ is preserved over $\mathcal{A}$, $\mathcal{W}$, and $\mathcal{J}$, we must show that the context remains consistent throughout the pipeline.

\begin{proof}
    The morphism $\mathcal{A}$ is defined as:
    \[
    \mathcal{A}: (E \times \Xi) \xrightarrow{\text{attest} \times \text{id}_{\Xi}} (C \times \Xi) \xrightarrow{\text{verify}} R \xrightarrow{\text{decide}} D
    \]
    The identity morphism $\text{id}_{\Xi}$ ensures that $\Xi$ is preserved through the $\text{attest}$ step. The $\text{verify}$ and $\text{decide}$ steps also operate over $C \times \Xi$ and $R$, respectively, without modifying $\Xi$. Thus, $\Xi$ is preserved in $\mathcal{A}$.
\end{proof}

\begin{proof}
    The morphism $\mathcal{W}$ is defined as:
    \[
    \mathcal{W} = \text{verify} \circ (\text{attest} \times \text{id}_{\Xi})
    \]
    The identity morphism $\text{id}_{\Xi}$ ensures that $\Xi$ is preserved through $\text{attest}$, and $\text{verify}$ operates over $C \times \Xi$ without altering $\Xi$. Thus, $\Xi$ is preserved in $\mathcal{W}$.
\end{proof}

\begin{proof}
    The morphism $\mathcal{J}$ is defined as:
    \[
    \mathcal{J} = \text{decide} \circ \text{verify}
    \]
    Since $\text{verify}$ preserves $\Xi$ and $\text{decide}$ operates on $R$ (which includes $\Xi$), $\Xi$ is preserved in $\mathcal{J}$.
\end{proof}

And further to this we additionally relate the preservation of the context $\Xi$ with $D$ as so:

\begin{proof}
    The identity morphism $\text{id}_{\Xi}$ ensures $\Xi$ remains invariant through $\mathcal{A}$, $\mathcal{W}$ and $\mathcal{J}$.

    Since $\Xi$ is preserved, the subobject classifier $D$ operates on a consistent context ensuring the integrity of the characteristic morphism $\mathcal{A}: E \to D$.

    The Heyting Algebra structure of $D$ relies on $\Xi$ for consistent evaluation of trust levels maintaining the logical properties of the topos.

    Therefore the preservation of $\Xi$ guarantees the integrity of the subobject classifier and the topos structure.
\end{proof}

Given this, in the event of a failure where $\mathcal{A}(e)\neq\top$, $\mathcal{W}$, and $\mathcal{J}$ can be used to provide us with the necessary \textit{forensics} to establish why the given element was not fully trusted \cite{oliver2020approach}. 

If $\mathcal{A}(e)\neq\bot$ and $\mathcal{W}(e,\xi)$ maps to $\text{im}(v_{ERR})$ providing us with knowledge that the failure was due to some procedural level, for example, an incorrect signature or some condition in $\Xi$ that led to that.

If $\bot<\mathcal{A}(e)\leq\top$ then  $\mathcal{W}(e,\xi)$ maps to non-error co-product in $R$ but $\mathcal{J}$ shows how this maps through to the actual decision. If $\mathcal{A}(e)=D_{AUTH}$ then we find through $\mathcal{J}$ that the decision has been made due to valid signature but no   measurements. Note the earlier distinction with having a valid signature but no measurements versus invalid measurements.

A variation of the above nuances in the policy, we can also use $\mathcal{J}$ to examine subtitles in the context, for example an element being `new' or not.

In conclusion $\mathcal{A}$ is providing us with the pipeline and $\mathcal{W}$, and $\mathcal{J}$ being able to give meaning to terms such as \textit{trustworthy} but also mechanisms to establish notions of forensics over this pipeline.

\subsection{Initial and Terminal Objects}\label{initialterminalobjectssection}
To ground our category we introduce initial and terminal objects as in figure \ref{initialterminalobjects}. The initial object is where the element has yet to come into existence and the terminal are when the system is out of existence.

\begin{figure}[ht]
\centering
\begin{tikzcd}[cramped]
	0 & E & C & R & D & {!}
	\arrow[from=1-1, to=1-2]
	\arrow["\cong"', curve={height=-30pt}, dotted, tail reversed, from=1-1, to=1-6]
	\arrow[from=1-2, to=1-3]
	\arrow[from=1-3, to=1-4]
	\arrow[from=1-4, to=1-5]
	\arrow[from=1-5, to=1-6]
\end{tikzcd}
\caption{Initial and Terminal Objects}\label{initialterminalobjects}
\end{figure}

There are two cases we can consider:

\begin{itemize}
 \item[] $O\rightarrow e_0 \rightarrow (\mathfrak{m_0}, \mathfrak{s_0})$
 \item[] $O\rightarrow e_1 \rightarrow (m_1, s_1)$
\end{itemize}

The first states that a system comes into existence but it not in a state where it can not provide attestation information - this might be seen in a pre-boot environment, or in some smaller microcontroller environments where such information. The definition of $C$ itself reinforces this $C \subseteq (M \times_E S) \backslash \{ ( \mathfrak{m_0}, \mathfrak{s_0})$. 

The second case is where the attestation can take place where $m_1$ \text{or} $s_1)$ could be null, but not both. This however suggests that there is more processing happening.

The terminal object is a `value' which denotes potentiality. In the Heyting Algebra we use for $D$, every element has a unique mapping $d\rightarrow !$, where $d\leq top$ meaning effectively that we do not have knowledge state of the system. As this is also true for for 0 and any unattestable object in $E$, we can assert that 0 and ! are isomorphic $0\cong !$. This characterises \otrust{} as a pointed category with a zero morphism $0_ED:E\rightarrow D$ which must factor through $\bot$ in the decision space ensuring that absence of information or non-existence is never classified erroneously as a higher trust level.

\subsection{Exponentials}\label{exponentials}
Now that we have the definitions and constraints of the overall attestation process defined we can turn to refining the morphisms $attest$, $verify$ and $decide$ and put these on a more concrete or functional basis. We also finally confirm that \otrust{} is a topos with the introduction of exponentials.

Earlier in eqn.\ref{decide} we defined an example $decide$ fold in terms of an explicit, total mapping from $R$ into $D$. The constraint being that $decide$ maps all co-products of $R$ into a point in the decision space. It is possible that two co-products of $R$ are mapped to the same \textit{decision} in $D$ (constraints on the algebra there permitting) and that some decision points are not mapped to. The exception of course being $R_{ERR}\rightarrow\bot$ and also the if $T$ is non-empty then something has to map to $\top$.

As there are potentially many decision mapping schemes, the choice of $decide$ is a point describing the fold in the space of all \textit{valid} decision mechanisms $\Delta:D^{R}$. For example eqn.\ref{altdecide} is equally a valid point in $\Delta$ as eqn.\ref{decide}.

 \begin{equation} 
 \begin{aligned} decide(r) = \begin{cases}
 \top  & \text{if } r \in im(v_1) \wedge \pi_{\Xi}(r) \\
 D_{AUTH}  & \text{if } r \in im(v_s) \wedge ameas(\pi_C(r))=0_M \\
 D_S & \text{if } r \in im(v_s) \wedge ameas(\pi_C(r)) \neq 0_M \\
 D_M  & \text{if } r \in im(v_m) \\
 \bot & \text{otherwise } r \in im(v_{ERR}) \\
 \end{cases}  \end{aligned} \label{altdecide}
 \end{equation}  

Similar treatment of $verify$ can be made resulting in a space $\Upsilon:R^C$ ensuring that the properties of this morphism as presented in eqn.\ref{verifymorphism} are also adhered as described earlier. 

Again continuing this $attest$ itself can be expressed in this form $\Alpha:C^E$ where the points in $\Alpha$ are the actual attestation mechanisms themselves. For example $\Alpha$ might contain familiar TPM operations such as \texttt{tpm\_quote} or  \texttt{tpm\_pcrread}, and also operations such as reading constructs such as the UEFI or IMA eventlogs, the machine ID parameter, or obtaining certain keys. 

A simple example might be executing:
\begin{verbatim}
tpm_quote(PCRs=`0,1,2,3', nonce=1234, sign_with=AK, ...)
\end{verbatim}
on some computer (with a TPM) producing a claim containing the measurements - the data part of the \texttt{TPMS\_ATTEST} structure and a signature of this with the attestation key of the TPM. If this were executed incorrectly or on a machine without a TPM then we would get a claim containing $(0_M,0_S)$ which by our earlier rules maps to $R_{ERR}$ and decision $\bot$ respectively.

An attestation server environment such as Keylime or Jane can now be expressed as an evaluation function that picks an element, requests the attestation data, verifies it, and then produces a decision. This is simply the process of implementing eqn.\ref{eval} and reporting if $e$ is \textit{fully trusted} when the characteristic morphism $\mathcal{A}$ holds.

\begin{equation}
eval:E\times(\Alpha, \Upsilon, \Delta) \rightarrow D
\label{eval}
\end{equation}

Interestingly, this also provides a comparison mechanism between attestation environments regarding the `size' of the spaces $\Alpha$, $\Upsilon$ and $\Delta$. The larger these are, the more fine-grained and/or flexible their capabilities are. A detailed discussion of this is outside the scope of this paper, but Keylime is tailored to TPM 2.0 based x86 UEFI machine, while Jane has mechanisms to handle much more fine-grained attestations, verifications and to a point, decisions too - along with much more complexity in operation however.
\subsection{Architectural Constriants}
The introduction of the exponentials $\Alpha$, $\Upsilon$ and $\Delta$ show in eqn. \ref{restricters} along with the evaluation function in eqn.\ref{eval} provide us with everything that a given system can do. However the selection of an attestation mechanism will restrict the verifications, ie: a TPM quote can only be verified by mechanisms that understand TPM quotes, and similarly certain results imply certain decision mechanisms and ultimately some attestation mechanisms might result in a restricted decision space.

\begin{equation} \begin{aligned} 
 \rho_\Alpha: & E\rightarrow\powerset{\Alpha}\\
 \rho_\Upsilon:& \Alpha\rightarrow\powerset{\Upsilon}\\
 \rho_\Delta:& \Upsilon\rightarrow\powerset{\Delta} 
 \end{aligned} 
 \label{restricters} \end{equation}

A the restriction $\rho_\Alpha$ picks which attestation functions can be sensibly applied to a given element. This then allows us to calculate how attestable and thus how potentially trustable a given element is. The trust potential is defined as in eqn.\ref{trustpotential}.

\begin{equation}
P(e)= \bigcup_{a\in\rho_\Alpha(e)} \left( \bigcup_{v\in\rho_\Upsilon(a)} \rho_\Delta(v) \right)
\label{trustpotential}
\end{equation}

An element consisting of a TPM, UEFI secure boot etc capabilities might admit many possible policies leading to decisions ranging fully from $\top$ to $\bot$ in $D$. We might say that these kinds of elements are \textit{highly attestable}.

Contrary to this an element that might only result in mappings to $\{D_M,\bot\}$ would reflect the limited attestability of that element, for example, a sensor that can only supply a non-cryptographically signed serial number.

This provides us with an interesting semantics to the term \textit{trustable}: 
\begin{itemize}
    \item an element where $P(e)=\{\bot\}$ is untrustable
    \item an element where $\{\top\}\in P(e)$ is potentially fully trustable
    \item given a lower bound $b$ on what we consider trustable, then if $\exists d\in P(e)\;\text{where}\;d\geq b$, then we can state that the element is only trustable with respect to that lower bound.
    \item similarly $\text{max}(P(e))$ provides the upper limit to how trustable an element could potentially be.
\end{itemize}

\section{Modelling Systems}
In this section we show examples of an attack, a method for extending the category to handle dynamic cases and its use in the specification of an attestation environment.

\subsection{Dynamicity: Booting, Run-time and Shutdown}
\begin{figure}[ht]
\centering
\includegraphics[scale=0.5]{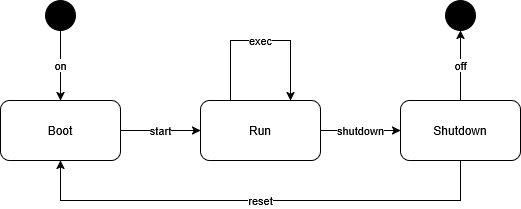}
\caption{Typical Boot-Run-Shutdown Sequence}
\label{pcsequence}
\end{figure}

In this section we create a simple model of a device such as an x86 PC with UEFI and describe how the system behaves over time. Consider the states in figure \ref{pcsequence} describing the lifecycle via $\sigma_X$ of some element $e$; the black circles denoting the initial and terminal objects.

We may assert a policy regarding trust levels as shown in eqn.\ref{trustlevels} to show that the boot sequence passes, say at least, secure boot tests, the runtime provides integrity and signature thorough measured boot backed up with TPM quotes, while the potential lowering of trust during shutdown corresponds with the termination of security services in the system. 

\begin{equation}
d_{AUTH}\leq\mathcal{A}(e_{boot})<\mathcal{A}(e_{run})\geq\mathcal{A}(e_{shutdown})
\label{trustlevels}
\end{equation}

In section \ref{initialterminalobjectssection} we introduced that $0\cong !$ which manifests in this example as off state of the system is isomorphic to the state before it is powered on, reinforcing the idea of the cyclic nature of the states of a machine. This however raises a complication when considering any saved or persistent state such as the TPM's registers which provide the source of the endorsement key. This is solved by our object $\Xi$ introduced earlier which describes some global context.

Having 0, 1 and $\Xi$ in our models now means that when the system transitions $e_{shutdown}\rightarrow !\cong0\rightarrow e1$ the element itself is reset, but the some state exists in $\Xi$. As stated a machine's (element's) identity is usually expressed as the endorsement key in the TPM chip; if the TPM is physically moved from one machine to another along with the NVRAM, then the new machine inherits the identity and measurements. In reality, the complexities of preserving measurements and physically moving the TPM and components containing the firmware and its NVRAM are exceptionally difficult; and even then if the measured boot includes, say, hardware measures (external cards, memory slot and chip identification etc) then these could never be successfully reconstructed\footnote{The Hardware Ship of Theseus. PCRs 2,3 and 6 are often a demonstration of such changes (in some systems). Some operating system manufacturers use similar ideas to detect and prevent their software being moved between machines, even after a simple hardware upgrade.}.

This also shows a distinction between $\sigma_{restart}$ and the power cycle $\sigma_{off}$ via $0\cong !$. The former can preserve information during the boot sequence. One example of this is the TPM's clock information structure which makes a distinction between restarting and resetting - the former monotonically increments a restart counter, the latter monotonically increments a reset counter and zeros the restart counter. 

This is used to track hibernations and suspensions, though strictly hibernation involves state being written to disk, in our case $\Xi$, and suspension additional state of $e$ and becomes another promising attack vector we can address here. We deliberately keep this model simple at this time however.

Two aspects arise from this model:
\begin{enumerate}
 \item what information is preserved across power cycles in $\Xi$
 \item what happens if something fails its attestation
\end{enumerate}

The former has been answered to a point, but directly addressing this allows reasoning about how contextual information affects the system and how that information needs to be taken into consideration during the verification and decision procedures.

The second point addresses the nature of the operations on the model and the constraints that these must work within to ensure that the system reaches suitable trust level as it moves to the next state. If attestation of the runtime does not achieve a trust level strictly greater than $d_{AUTH}$ then we break the statement in eqn.\ref{trustlevels}.

This also leads us to the nature of the operations and whether these can be further classified into operations that might potentially change the trust levels of the system. If we model the operations as exponents and then classify these as in eqn.\ref{sigmaclassification} then we start to obtain visibility into the deeper nature of these operations and what might be permissible in given situations.

\begin{equation}
\Sigma:E^E\;\text{where}\;\Sigma\cong\coprod_{s\in\{\text{idempotent},\text{dangerous},\ldots\}} s
\label{sigmaclassification}
\end{equation}

Where \textit{idempotent} operations must preserve the trust-level as seen in eqn.\ref{idempotent}, \textit{dangerous} operations may lower it as shown in eqn.\ref{dangerous}.

\begin{equation}
\forall e:E,\sigma:\Sigma_{idempotent}\bullet \mathcal{A}(e)=\mathcal{A}(\sigma(e))
\label{idempotent}
\end{equation}
 
\begin{equation}
\forall e:E,\sigma:\Sigma_{dangerous}\bullet \mathcal{A}(e)\geq\mathcal{A}(\sigma(e))
\label{dangerous}
\end{equation}

This takes us also to a philosophical discuss about what happens if an operation [at runtime] can potentially \textit{increase} the trust level, even if idempotent or dangerous or otherwise. From a security point of view, dangerous operations can be restricted or only applied under certain circumstances.

What this model gives us is the language to clearly state what is required of a system in terms of its behaviour and how that should be specified and understood. The weakness here is that our systems are treated as individual points or black boxes, but for the moment this is sufficient without adding unnecessary complexity to the discussion.

\subsection{Evil Maid}
The Evil Maid Attack\footnote{The name is attributed to Joanna Rutkowska who coined it in 2009} \cite{knapp2024should} can be modelled using \otrust{}. The attack occurs during the sensitive transition from potentiality to operation. The Evil Maid replaces the legitimate firmware $f$ with a malicious version $f'$ while the system is powered off \cite{bashun2013too,ruytenberg2022lightning}. Effectively $f'$ is a failure of the pullback of the subobject classifier $\chi_\top$.

We are now considering $E$ to be a more complex structure $E\cong Firmware\times Hardware\times Software$ \cite{legatiuk2017categorical}, equipped with projection morphisms such as $\pi_{firmware}:E\rightarrow Firmware$; let $f=\pi_{firmware}(e)$ below:

When the system is powered on, the sequence proceeds:
\begin{itemize}
    \item Power on $0\xrightarrow{\sigma_{on}}e1$ with the identity being established in the TPM in context $\Xi$
    \item Instead of $\sigma_{boot}$, we execute $\sigma_{boot'}$ leading to $e2'$
    \item Now at $e2'$ the system consists of mutated firmware $f'$ and identity from $\Xi$
\end{itemize}

We now apply the characteristic morphism $\mathcal{A}$ to the state $e2'$:

\begin{itemize}
  \item Identity verification via $S$ succeeds, therefore for the identity part, the decision $d\geq D_{AUTH}$
  \item Integrity verification via $M$ measures $p_f'$ and since $p_f'\not\in T$ (the trusted subobject), the verify morphism maps this claim to the error object $R_{ERR}$ (the failure state).
  \item The decision via $decide$ must map $R_{ERR}$ to $\bot\in D$
  \item $D_{AUTH} \wedge \bot = \bot$
\end{itemize}

\subsection{Attestation Server Design}
A number of attestation server environments have been developed such as OpenAttestation and RedHat's Keylime, and which are generally available, as well as internal systems utilised by Microsoft Azure and Google. We however have developed for research purposes over the past decade a system now named \textit{Jane} - formerly Nokia Attestation Environment\footnote{available at: \url{https://github.com/iolivergithub/jane}}. The premise of Jane was to implement attestation functions at a very fine-grained level and explicitly \textbf{not} to enforce decisions such as trusted or untrusted explicitly. 

\begin{figure}[ht]
\centering
\includegraphics[scale=0.34]{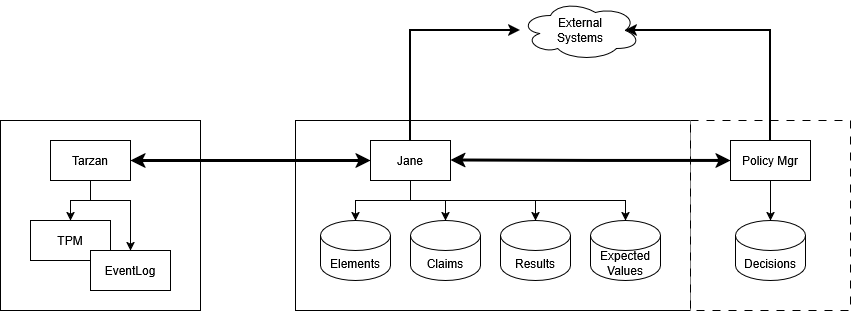}
\caption{Jane Architecture}\label{janearchitecture}
\end{figure}

Figure \ref{janearchitecture} shows the overall architecture with the \textit{Tarzan} trust agent (Jane can happily work with such as Gramine and CoCo for confidential containers, Keylime, Veraison RATSD\footnote{\url{https://github.com/veraison/ratsd}} and others as necessary) as well as any internal source of measurements and signatures, ie: TPM, UEFI, T2 etc. The data model utilised follows the model in figures \ref{attestationpipeline} and \ref{pullbackclaims}.

While the usual software engineering compromises have had to have been made, selection of implementation language, database structures and the semantics surrounding this\footnote{The authors wish to note that Haskell would have been a good idea with hindsight}, the overall functionality and design very much mirrors the original category theoretic specification. In particular the explicit use of exponentiation in section \ref{exponentials} allowing us to abstract and template rules, attestation mechanisms and communication protocols.

\begin{figure}[ht]
\centering
\includegraphics[scale=0.2]{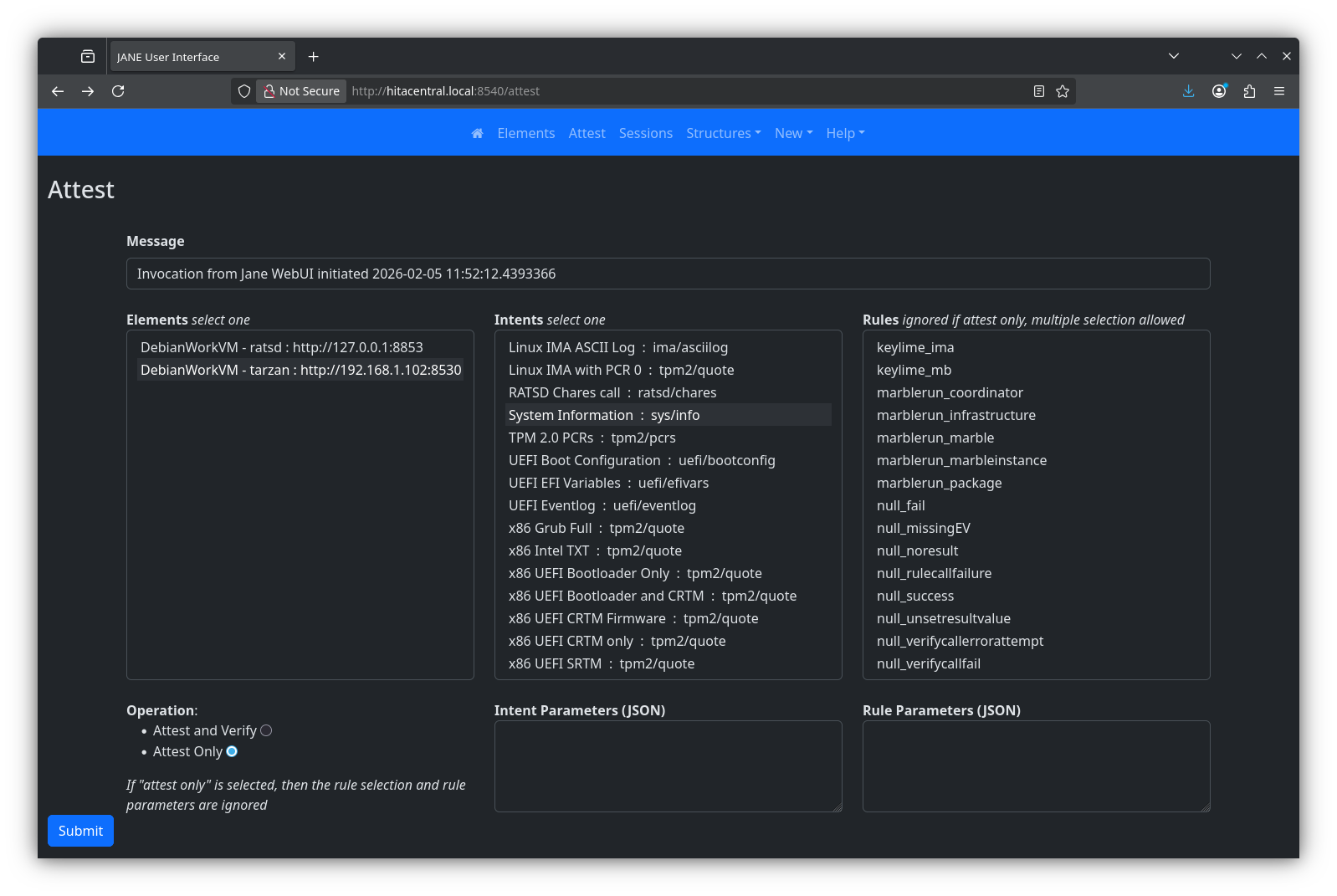}
\caption{Attestation Functionality UI in Jane}\label{attest}
\end{figure}

Figure \ref{attest} shows the attestation functionality UI in Jane. Each intent and rule being effectively being objects $\Alpha$ and $\Upsilon$, and in this case the two attestation options being implemented as in eqn.\ref{janeattestoptions}. The identifier $I$ and the external information $\Xi$ being provided either by the user and/or by Jane.

\begin{equation}
\begin{aligned}
eval_{AO} : & E\times \Alpha\times I \times\Xi \rightarrow C \\
eval_{AV} : & eval_{AO} \times \Upsilon\rightarrow R 
\end{aligned}
\label{janeattestoptions}
\end{equation}

The explicit split of decision away from attestation and verification allows us much more freedom in both \textit{how} to decide whether a system is trusted, and also to enforce much more localised policies as we have seen in larger telecoms deployments where this has been tested. A specific decision/policy component has been developed as well as proof-of-concept integrations with a number of system management tools for ``mass-attestation" and trust decisions across a large system. The use of Jane for digital forensics is discussed in \cite{oliver2020approach}.

\section{Systems and Compositions}
In this section we outline some ideas regarding how systems are composed, attested and what trust means in these cases. Elements may be composed of other elements and we model this through an endomorphism $\kappa:E\rightarrow E$ where the structure forms a transitive closure given any element in that structure.

For example, we might have a simple element $e1$ composed of two `sub'-elements $e2$ and $e3$. Each of these elements individually can be attested and that the overall trust of the system given in eqn.\ref{minimalsystemtrust} is the minimum of all the trust levels of the components of the system:

\begin{equation}
 \mathcal{A}^+(e1)=\text{min}(\{\mathcal{A}(e1),\mathcal{A}(e2),\mathcal{A}(e3),\ldots\})
 \label{minimalsystemtrust}
\end{equation}

While this might be a reasonable and logical interpretation, in reality however the trust in a system does not appear in this manner and that internal components are not visible for individual attestation. Indeed this is one of they key points raised in Luhmann's systems theory and trust where the trust of a system is only `attestable' at its outermost point, that of the system boundary itself. The system may `protect' its internals \cite{HOLMSTROM2007255} which is typically seen in real-life systems.

\begin{figure}[ht] 
 \centering 
 \begin{tabular}{cc}
 \includegraphics[scale=0.4]{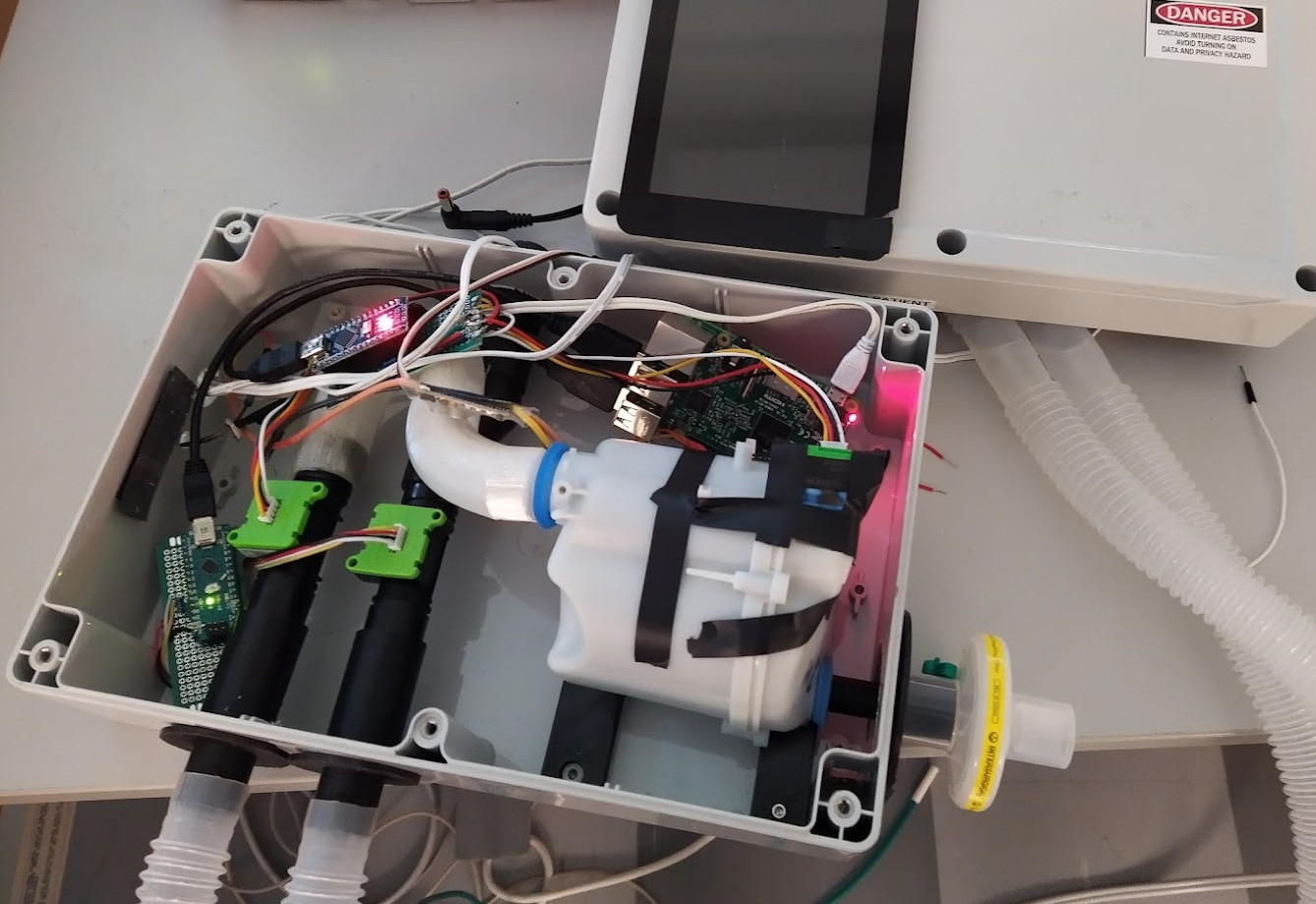} &
 \includegraphics[scale=0.32]{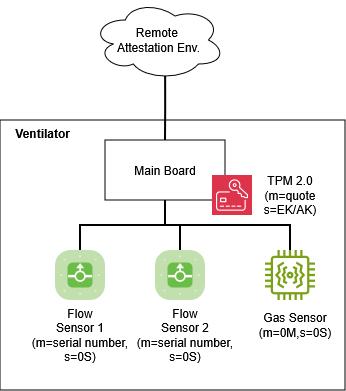}
 \end{tabular}
 \caption{Example Medical Device (Ventilator)}
 \label{medicaldevice} 
 \end{figure}

Consider a medical device shown in figure \ref{medicaldevice} whose primary interface board has a TPM but which also has sensors internally that can provide serial numbers but no cryptographic proof of this. Attesting the system as a whole might result in a trusted decision; internally the attestation procedures on the sensors themselves could either be hidden (and different, $D$ would be simply $\{\top,\bot\}$ and/or some information about this presented upwards to the composing element \cite{oliver2020approach}.

In this sense the system composition morphism $\kappa$ isn't just a statement that one element is composed of one or more others, but that it mediates the trust decision. Rather than $\mathcal{A}(mainboard)=\bot$ because $\mathcal{A}(gassensor)=\bot$, we may have $\mathcal{A}\circ\kappa(gassensor)=\top$ but with the mainboard mediating the trust to $\top$ because it can `protect' the system integrity in some way. In effect this creates local trust logics which are not easily handled by the \otrust{} model.

This demonstrates that composition in a system is a more complex process than just stating that elements are simply connected to each other. Indeed if we expand out our definition of $\kappa$ to include different kinds of composition and communication then these localised trust logics become more important. 

There are a number of mechanisms available to use to better express these properties: $\kappa$ as a functor $E\rightarrow E$ in some category of elements $\mathcal{E}$ etc. The idea of a monoidal category $(\mathcal{E},\otimes,\mathcal{I})$ appears not to be sufficient as $\mathcal{A}(e1\otimes e2)=\mathcal{A}(e1)\wedge\mathcal{A}(e2)$ implying our weakest trust model in eqn.\ref{minimalsystemtrust}. The use of Kan-extensions and a move to a 2-category are being considered, and all of this is considered future work.

\section{Conclusions}
We have presented a category \otrust{} to model the notion of trust and attestation as applied to what are termed `trusted systems'. Properties such as how the process is decomposed into the attestation, verification and decision parts, as well as additional structures was discussed. Further properties such as the initial and terminal objects, and notions of trustworthiness were introduced. This was then expanded to describe how attestation, verification and decision operations could be composed and the constraints over these along with a notion of a metric pertaining to the expressibility of the trust properties of a system.

Demonstrating this we included a simplified example of a typical machine with boot, run-time and shutdown operations, an attack such as Evil Maid and finally a description of how this model is used to specify the workings of a remote attestation environment.

\otrust{} is a static logic which is used to reason about states but lacks the mechanics to reason about the evolution of rules or policy-negotiation. Despite this our category is sufficient for specification and exploration of terminology, semantics and aspects such as result types and decision mechanisms and classifications. This provides a clear framework in which the concepts surrounding system trust, integrity and attestation can be discussed unambiguously. As noted in the worked examples, it has provided to be an invaluable tool.

As more structure is added to \otrust{} the more complex the model becomes; this is certainly the case when we started expressing state transitions and compositionality in $E$ with further endomorphisms. We also avoided topics such as recovery of systems from lesser trusted stated, though the morphism $\mathcal{J}$ as its use in digital forensics provides interesting possibilities for the formalisation of this.

Options such as functors, Kan extensions and categorification \cite{baez1998categorification} were proposed but these require further work. However it serves to demonstrate the complexity of attesting trust in both dynamic and composed systems.

Another consideration for the model is how it interacts with other categories - one of the main features of category theory is composition. Two areas are of interest: risk and actions, and then linking these possibly back to the $\sigma$ operations on elements to restore trust in a system. Category theory has been utilised in certain domains for specifying action models \cite{garrett2020application} based on \textit{Observe, Orient, Decide, Act (OODA)} loops: $E\xrightarrow[]{\mathcal{A}} D\rightarrow ACT$ where Observe and Orient are congruent to $\mathcal{W}$.

As final note, and as paranoid cybersecurity researchers we agree with the premise that the best recovery mechanism for a firmware compromised machine involves copious amounts of petrol and matches which may be formalised as the \textit{unique terminal endomorphism} $fire:*\rightarrow !$.

\bibliographystyle{plain}
\bibliography{refs} 

@article{Lipner2015TheBA,
  title={The Birth and Death of the Orange Book},
  author={Steven B. Lipner},
  journal={IEEE Annals of the History of Computing},
  year={2015},
  volume={37},
  pages={19-31},
  url={https://api.semanticscholar.org/CorpusID:16625319}
}

@ARTICLE{11078841,
  author={Fettweis, Gerhard P. and Grünberg, Patricia and Hentschel, Tim and Köpsell, Stefan},
  journal={IEEE Communications Magazine}, 
  title={Conceptualizing Trustworthiness and Trust in Communications}, 
  year={2025},
  volume={63},
  number={12},
  pages={126-132},
  doi={10.1109/MCOM.001.2400383}
}

@article{kumar2020evolution,
  title={The evolution of trust and trustworthiness},
  author={Kumar, Aanjaneya and Capraro, Valerio and Perc, Matja{\v{z}}},
  journal={Journal of the Royal Society Interface},
  volume={17},
  number={169},
  pages={20200491},
  year={2020},
  publisher={The Royal Society}
}

@book{hardin2002trust,
  title={Trust and trustworthiness},
  author={Hardin, Russell},
  year={2002},
  publisher={Russell Sage Foundation}
}

@article{HOLMSTROM2007255,
title = {Niklas Luhmann: Contingency, risk, trust and reflection},
journal = {Public Relations Review},
volume = {33},
number = {3},
pages = {255-262},
year = {2007},
note = {Special Issue on Social Theory},
issn = {0363-8111},
doi = {https://doi.org/10.1016/j.pubrev.2007.05.003},
url = {https://www.sciencedirect.com/science/article/pii/S0363811107000574},
author = {Susanne Holmstr\"om}
}

@book{booth20225g,
  title={5G hardware supply chain security through physical measurements},
  author={Booth, James C and Booth, James C and Dowell, Marla L and Feldman, Ari D and Hale, Paul D and Midzor, Melissa M and Orloff, Nathan D},
  year={2022},
  publisher={US Department of Commerce, National Institute of Standards and Technology}
}

@article{cherkaoui2025categorical,
  title={Categorical Framework for Quantum-Resistant Zero-Trust AI Security},
  author={Cherkaoui, I and Clarke, C and Horgan, J and Dey, I},
  journal={arXiv preprint arXiv:2511.21768},
  year={2025}
}

@inproceedings{datta2009logic,
  title={A logic of secure systems and its application to trusted computing},
  author={Datta, Anupam and Franklin, Jason and Garg, Deepak and Kaynar, Dilsun},
  booktitle={2009 30th IEEE Symposium on Security and Privacy},
  pages={221--236},
  year={2009},
  organization={IEEE}
}

@inproceedings{legatiuk2017categorical,
  title={A categorical approach towards metamodeling cyber-physical systems},
  author={Legatiuk, Dmitrii and Theiler, Michael and Dragos, Kosmas and Smarsly, Kay},
  booktitle={Proceedings of the 11th International Workshop on Structural Health Monitoring},
  volume={9(12)},
  pages={1--8},
  year={2017}
}

@inproceedings{ren2025algebraic,
  title={An Algebraic Approach to Asymmetric Delegation and Polymorphic Label Inference},
  author={Ren, Silei and Acay, Co{\c{s}}ku and Myers, Andrew C},
  booktitle={European Symposium on Research in Computer Security},
  pages={334--353},
  year={2025},
  organization={Springer}
}

@inproceedings{oliver2020approach,
  title={An approach to combining medical device fault analysis with trusted computing forensics},
  author={Oliver, Ian},
  booktitle={2020 IEEE International Conference on Big Data (Big Data)},
  pages={1831--1837},
  year={2020},
  organization={IEEE}
}

@article{amoore2014security,
  title={Security and the incalculable},
  author={Amoore, Louise},
  journal={Security Dialogue},
  volume={45},
  number={5},
  pages={423--439},
  year={2014},
  publisher={SAGE Publications Sage UK: London, England}
}

@article{garrett2020application,
  title={The Application of Graph Theory to Modeling, Simulation, Analysis Looping and Trust to Quantify Mission Success},
  author={Garrett Jr, Robert K and Fairbanks, James P and Loper, Margaret L and Moreland Jr, James D},
  year={2020},
  publisher={Defense Technical Information Center (DTIC)},
  notes={https://apps.dtic.mil/sti/trecms/pdf/AD1125185.pdf}
}

@inproceedings{oliver2009proposed,
  title={A proposed diagrammatic logic for ontology specification and visualization},
  author={Oliver, Ian and Howse, John and Stapleton, Gem and Nuutila, Esko and T{\"o}rma, S},
  booktitle={International Semantic Web Conference},
  year={2009}
}

@INPROCEEDINGS{8401590,
  author={Oliver, Ian and Panda, Sakshyam and Wang, Ke and Kalliola, Aapo},
  booktitle={2018 21st Conference on Innovation in Clouds, Internet and Networks and Workshops (ICIN)}, 
  title={Modelling NFV concepts with ontologies}, 
  year={2018},
  volume={},
  number={},
  pages={1-7},
  doi={10.1109/ICIN.2018.8401590}}

@book{fong2019invitation,
  title={An invitation to applied category theory: seven sketches in compositionality},
  author={Fong, Brendan and Spivak, David I},
  year={2019},
  publisher={Cambridge University Press}
}

@phdthesis{ruytenberg2022lightning,
  title={When lightning strikes thrice: breaking Thunderbolt security},
  author={Ruytenberg, Bj{\"o}rn},
  year={2022},
  school={Master’s thesis, Eindhoven University of Technology}
}

@inproceedings{bashun2013too,
  title={Too young to be secure: Analysis of UEFI threats and vulnerabilities},
  author={Bashun, Vladimir and Sergeev, Anton and Minchenkov, Victor and Yakovlev, Alexandr},
  booktitle={14th Conference of Open Innovation Association FRUCT},
  pages={16--24},
  year={2013},
  organization={IEEE}
}

@inproceedings{knapp2024should,
  title={Should Smart Homes Be Afraid of Evil Maids?: Identifying Vulnerabilities in IoT Device Firmware},
  author={Knapp, Austen and Wamuo, Emmanuel and Rahat, Minhajul Alam and Torres-Arias, Santiago and Bloom, Gedare and Zhuang, Yanyan},
  booktitle={2024 IEEE 14th Annual Computing and Communication Workshop and Conference (CCWC)},
  pages={0467--0473},
  year={2024},
  organization={IEEE}
}

@article{baez1998categorification,
  title={Categorification},
  author={Baez, John C and Dolan, James},
  journal={arXiv preprint math/9802029},
  year={1998}
}

@inproceedings{10.1145/3567445.3571105,
author = {Bradbury, Matthew and Prince, Daniel and Marcinkiewicz, Victoria and Watson, Tim},
title = {Attributes and Dimensions of Trust in Secure Systems},
year = {2023},
isbn = {9781450396653},
publisher = {Association for Computing Machinery},
address = {New York, NY, USA},
url = {https://doi.org/10.1145/3567445.3571105},
doi = {10.1145/3567445.3571105},
booktitle = {Proceedings of the 12th International Conference on the Internet of Things},
pages = {179–186},
numpages = {8},
location = {Delft, Netherlands},
series = {IoT '22}
}

@book{goldblatt2014topoi,
  title={Topoi: the categorial analysis of logic},
  author={Goldblatt, Robert},
  volume={98},
  year={2014},
  publisher={Elsevier}
}

@inproceedings{coker2008attestation,
  title={Attestation: Evidence and trust},
  author={Coker, George and Guttman, Joshua and Loscocco, Peter and Sheehy, Justin and Sniffen, Brian},
  booktitle={International Conference on Information and Communications Security},
  pages={1--18},
  year={2008},
  organization={Springer}
}

@InProceedings{10.1007/978-3-540-45217-1_18,
author="Golbeck, Jennifer
and Parsia, Bijan
and Hendler, James",
editor="Klusch, Matthias
and Omicini, Andrea
and Ossowski, Sascha
and Laamanen, Heimo",
title="Trust Networks on the Semantic Web",
booktitle="Cooperative Information Agents VII",
year="2003",
publisher="Springer Berlin Heidelberg",
address="Berlin, Heidelberg",
pages="238--249",
isbn="978-3-540-45217-1"
}

@InProceedings{10.1007/978-3-030-23182-8_10,
author="Oliver, Ian
and Howse, John
and Stapleton, Gem
and Shams, Zohreh
and Jamnik, Mateja",
editor="Endres, Dominik
and Alam, Mehwish
and {\c{S}}otropa, Diana",
title="Exploring and Conceptualising Attestation",
booktitle="Graph-Based Representation and Reasoning",
year="2019",
publisher="Springer International Publishing",
address="Cham",
pages="131--145",
isbn="978-3-030-23182-8"
}

@book{spivak2014category,
  title={Category theory for the sciences},
  author={Spivak, David I},
  year={2014},
  publisher={MIT press}
}

@inproceedings{shaikh2012trust,
  title={Trust framework for calculating security strength of a cloud service},
  author={Shaikh, Rizwana and Sasikumar, M},
  booktitle={2012 International Conference on Communication, Information \& Computing Technology (ICCICT)},
  pages={1--6},
  year={2012},
  organization={IEEE}
}

@article{ali2022maturity,
  title={A maturity framework for zero-trust security in multiaccess edge computing},
  author={Ali, Belal and Hijjawi, Simsam and Campbell, Leith H and Gregory, Mark A and Li, Shuo},
  journal={Security and Communication Networks},
  volume={2022},
  number={1},
  pages={3178760},
  year={2022},
  publisher={Wiley Online Library}
}

@article{yan2016security,
  title={A security and trust framework for virtualized networks and software-defined networking},
  author={Yan, Zheng and Zhang, Peng and Vasilakos, Athanasios V},
  journal={Security and communication networks},
  volume={9},
  number={16},
  pages={3059--3069},
  year={2016},
  publisher={Wiley Online Library}
}

@article{primiero2017trust,
  title={Trust and distrust in contradictory information transmission},
  author={Primiero, Giuseppe and Raimondi, Franco and Bottone, Michele and Tagliabue, Jacopo},
  journal={Applied Network Science},
  volume={2},
  number={1},
  pages={12},
  year={2017},
  publisher={Springer}
}

@INPROCEEDINGS{7092916,
  author={Berger, Stefan and Goldman, Kenneth and Pendarakis, Dimitrios and Safford, David and Valdez, Enriquillo and Zohar, Mimi},
  booktitle={2015 IEEE International Conference on Cloud Engineering}, 
  title={Scalable Attestation: A Step Toward Secure and Trusted Clouds}, 
  year={2015},
  volume={},
  number={},
  pages={185-194},
  doi={10.1109/IC2E.2015.32}}

@inproceedings{joque2020deconstructing,
  title={Deconstructing cybersecurity: From ontological security to ontological insecurity},
  author={Joque, Justin and Haque, SM Taiabul},
  booktitle={Proceedings of the New Security Paradigms Workshop 2020},
  pages={99--110},
  year={2020}
}

@article{mabrok2017category,
  title={Category theory as a formal mathematical foundation for model-based systems engineering},
  author={Mabrok, Mohamed A and Ryan, Michael J},
  journal={Appl. Math. Inf. Sci},
  volume={11},
  number={1},
  pages={43--51},
  year={2017}
}

@article{diskin2014category,
  title={Category theory and model-driven engineering: From formal semantics to design patterns and beyond},
  author={Diskin, Zinovy and Maibaum, Tom},
  journal={Model-Driven Engineering of Information Systems: Principles, Techniques, and Practice},
  volume={173},
  year={2014},
  publisher={Apple Academic Press}
}

@inproceedings{guha2004propagation,
  title={Propagation of trust and distrust},
  author={Guha, Ramanthan and Kumar, Ravi and Raghavan, Prabhakar and Tomkins, Andrew},
  booktitle={Proceedings of the 13th international conference on World Wide Web},
  pages={403--412},
  year={2004}
}

@misc{rfc9334,
    series =    {Request for Comments},
    number =    9334,
    howpublished =  {RFC 9334},
    publisher = {RFC Editor},
    doi =       {10.17487/RFC9334},
    url =       {https://www.rfc-editor.org/info/rfc9334},
    author =    {Henk Birkholz and Dave Thaler and Michael Richardson and Ned Smith and Wei Pan},
    title =     {{Remote ATtestation procedureS (RATS) Architecture}},
    pagetotal = 46,
    year =      2023,
    month =     jan
}

@techreport{nibaldi1979proposed,
  title={Proposed Technical Evaluation Criteria for Trusted Computer Systems},
  institution={MITRE},
  author={Nibaldi, GH},
  year={1979}
}

@inproceedings{ferro2024standard,
  title={Standard-Based Remote Attestation: The Veraison Project},
  author={Ferro, Lorenzo and Lioy, Antonio and others},
  booktitle={Proceedings of the Italian Conference on Cybersecurity (ITASEC 2024), CEUR-WS, Salerno, Italy},
  pages={9--11},
  year={2024}
}

@inproceedings{ott2023universal,
  title={Universal remote attestation for cloud and edge platforms},
  author={Ott, Simon and Kamhuber, Monika and Pecholt, Joana and Wessel, Sascha},
  booktitle={Proceedings of the 18th International Conference on Availability, Reliability and Security},
  pages={1--11},
  year={2023}
}

@article{thompson1984reflections,
  title={Reflections on trusting trust},
  author={Thompson, Ken},
  journal={Communications of the ACM},
  volume={27},
  number={8},
  pages={761--763},
  year={1984},
  publisher={ACM New York, NY, USA}
}

@article{gambetta2000can,
  title={Can we trust trust},
  author={Gambetta, Diego and others},
  journal={Trust: Making and breaking cooperative relations},
  volume={13},
  number={1},
  pages={213--237},
  year={2000}
}

@book{luhmann2013introduction,
  title={Introduction to systems theory},
  author={Luhmann, Niklas and Baecker, Dirk and Gilgen, Peter},
  year={2013},
  publisher={Polity Cambridge}
}

@book{arthur2015practical,
  title={A practical guide to TPM 2.0: Using the new trusted platform module in the new age of security},
  author={Arthur, Will and Challener, David and Goldman, Kenneth},
  year={2015},
  publisher={Springer Nature}
}

@article{moschovakis1999intuitionistic,
  title={Intuitionistic logic},
  author={Moschovakis, Joan},
  journal={Stanford Encyclopedia of Philosophy},
  year={1999}
}

@incollection{kripke1965semantical,
  title={Semantical analysis of intuitionistic logic I},
  author={Kripke, Saul A},
  booktitle={Studies in Logic and the Foundations of Mathematics},
  volume={40},
  pages={92--130},
  year={1965},
  publisher={Elsevier}
}

@inproceedings{becker2012foundations,
  title={Foundations of logic-based trust management},
  author={Becker, Moritz Y and Russo, Alessandra and Sultana, Nik},
  booktitle={2012 IEEE Symposium on Security and Privacy},
  pages={161--175},
  year={2012},
  organization={IEEE}
}

@book{simpson2023trust,
  title={Trust: A philosophical study},
  author={Simpson, Thomas W},
  year={2023},
  publisher={Oxford University Press}
}


\end{document}